\documentclass[aps,prb,twocolumn,floatfix,showpacs,superscriptaddress,10pt,noshowpacs]{revtex4}
\pdfoutput=1 
\usepackage[english]{babel}
\usepackage{amsmath,amssymb}
\usepackage{graphicx}
\usepackage{color}
\usepackage{bm}
\usepackage{amssymb}
\usepackage{float}

\begin{document}
\title{Comparison of the dislocation density obtained by HR-EBSD and X-ray profile analysis}

\author{Szilvia Kal\'acska}
\affiliation{Department of Materials Physics, E\"otv\"os University Budapest,
H-1517 Budapest POB 32, Hungary}

\author{Istv\'an Groma}
\email{groma@metal.elte.hu}
\affiliation{Department of Materials Physics, E\"otv\"os University Budapest,
H-1517 Budapest POB 32, Hungary}

\author{Andr\'as Borb\'ely}
\affiliation{Ecole Nationale Sup\'erieure des Mines, SMS-EMSE, CNRS:UMR 5307, LGF, 42023, 
Saint-Etienne Cedex 2, France}

\author{P\'eter Dus\'an Isp\'anovity}
\affiliation{Department of Materials Physics, E\"otv\"os University Budapest,
H-1517 Budapest POB 32, Hungary}

\begin{abstract}
Based on the cross correlation analysis of the Kikuchi diffraction patterns high-resolution EBSD is 
a well established method to determine the internal stress in deformed crystalline materials. In 
many cases, however, the stress values obtained at the different scanning points have a large (in 
the order of GPa) scatter. As it was first demonstrated by Wilkinson and co-workers this is due to 
the long tail of the probability distribution of the internal 
stress ($P(\sigma)$) generated by the dislocations present in the system. According to the 
theoretical investigations of Groma and co-workers the tail of $P(\sigma)$ is inverse cubic with 
prefactor proportional to the total dislocation density  $<\rho>$. In this paper we 
present a direct comparison of the X-ray line broadening and $P(\sigma)$ obtained by EBSD on 
deformed Cu single crystals.  It is shown that $<\rho>$ can be determined from $P(\sigma)$. This 
opens new perspectives for the application of EBSD in determining mesoscale parameters in a 
heterogeneous sample.  
\end{abstract}

\date{\today}

\maketitle

Quantitative characterization of plastically deformed crystals in terms of dislocation density by 
transmission electron microscopy (TEM) and 
X-ray diffraction was a very important step in the development of basic models of crystal plasticity \cite{mughrabi2002}. 
This is especially true in the case of the composite model \cite{mughrabi} of heterogeneous 
dislocation structures, which postulates Taylor type \cite{Taylor1934} relations between the local 
flow stress and local dislocation density. 
Accessing local field quantities, however, requires methods capable to capture structural 
heterogeneities 
at the sub-micrometer scale, which can be done with the TEM, but obtaining statistically significant information  
necessitates a large amount of work. Development of novel automated characterization methods providing dislocation density 
data at the local scale is therefore important.

The aim of the present work is to present a new method for the evaluation of the average 
dislocation density at the mesoscale. 
It is based on the statistical properties of the distribution of local stresses determined by 
high-resolution backscatter electron diffraction (HR-EBSD)\cite{wilkinson2010}. 
To address their physical significance, the results will be compared to the outcome of discrete dislocation dynamics 
simulations and X-ray diffraction (XRD) line profile analysis\cite{groma1998x}. 
The latter is a well established experimental technique for determining microstructural parameters such as 
coherent domain size, dislocation density and its fluctuation. 
As shown by Groma {\it et. al.}\cite{groma1998x, groma2000,borbely2001,groma2013} in the so-called 
"strain broadening" 
setup \cite{groma1998x} the two leading terms of the asymptotic decay region of the intensity distribution $I(q)$ read as 
\begin{eqnarray}
            I(q)=\frac{1}{\pi^2 d}\frac{1}{q^2}+\frac{\Lambda}{4\pi^2}<\rho>\frac{1}{q^3} \ \ 
|q|>q_0
\label{eq:I}
\end{eqnarray}
where $q=2[\sin(\Theta)-\sin(\Theta_0)]/\lambda$,
 $d$ is the coherent domain size, $<\rho>$ is the average dislocation density and $\lambda$ is 
the wavelength of the X-rays. $\Theta$ and $\Theta_0$ are the half of the scattering angle and the Bragg 
angle, respectively. The parameter $\Lambda$ is commonly given in the form 
$\Lambda=2|\vec{g}|^2|\vec{b}|^2C_g/\pi$ where $\vec{b}$ and $\vec{g}$ are the Burgers and
the diffraction vector, respectively. $C_g$ is called the {\it diffraction contrast factor} and depends on the type of the dislocation 
and the relative geometrical position between the dislocation line direction and the direction of $\vec{g}$. 
A detailed description of the contrast factor calculation can be found in \cite{borbely2001}. 
A remarkable feature of Eq.~(\ref{eq:I}) is its independence from the configuration of 
dislocations usually described 
in terms of dislocation-dislocation correlations. Certainly, the $q_0$ value from which 
Eq.~(\ref{eq:I}) describes well the 
asymptotic region depends on correlations as it will be exemplified later. 
Considering that the tail of the experimental intensity curve can be rather noisy the actual values of the 
domain size and the dislocation density can be better obtained from the integral quantity 
\begin{eqnarray}
            M_2(q)=\int_{-q}^q q'^2 I(q') dq' \label{eq:M_2}
\end{eqnarray}
called as second order restricted moment \cite{groma1998x,borbely2001}. Analyzing higher order 
restricted moments can also be useful \cite{groma2000,borbely2001}, 
but for the experimental investigations presented here the use of $M_2(q)$ is enough. 
After substituting Eq.~(\ref{eq:I}) into Eq.~(\ref{eq:M_2}) at large enough $q$ values we get
\begin{eqnarray}
   M_2(q)=\frac{1}{\pi^2 d}q+\frac{\Lambda}{2\pi^2}<\rho>\ln\left(\frac{q}{q_0}\right ) 
\label{eq:M_2_rho} 
\end{eqnarray}
where $q_0$ is a constant depending on the dislocation-dislocation correlation. If the coherent 
domain size is larger than of about 1 $\mu$m the first term in Eq.~(\ref{eq:M_2_rho}) becomes 
negligible beside the
contribution of dislocations and the plot of $M_2$ versus $\ln(q)$ becomes a straight line 
in the asymptotic regime $q \rightarrow \infty$. Its slope is proportional to the mean
dislocation density. Using this feature the dislocation density can be determined with an accuracy of a few percent.  

HR-EBSD is a scanning electron microscope (SEM) based method, which allows determining the 
stress/strain in a crystalline material at the length scale of tens of nanometers. 
It is based on a cross-correlation method 
\cite{wilkinson2006,wilkinson2010,wilkinson2011,britton2012,jiang2013,jiang2015,wilkinson2014} exploiting
small changes in backscattered Kikuchi diffraction patterns corresponding to reference point and the 
actual point analysed. 
A detailed description of the technique can be found in \cite{wilkinson2006,britton2012}. It was first demonstrated by 
Wilkinson {\it et al.} \cite{jiang2013,jiang2015,wilkinson2014} that local stress values in 
deformed polycrystal can be unexpectedly high and vary by much as $\pm$1 GPa. This 
unusual behavior is the consequence of the $1/r$ type long-range stress field generated 
by a dislocation. According to the analytical calculations of Groma {\it et al.} 
\cite{groma1998,groma2004} the tail of the probability distribution density of the internal stress generated by 
a set of straight parallel dislocations decays as 
\begin{eqnarray}
 P(\sigma)\rightarrow G^2 b^2 C_{\sigma} <\rho>\frac{1}{\sigma^3} \label{eq:P} 
\end{eqnarray}
where $G$ is the shear modulus and $C_{\sigma}$ (in analogy with XRD) could be called as the {\it 
stress contrast factor}
since its value depends on the type of dislocation, its line direction and the stress component under consideration \cite{groma1998}.
Similarly to the X-ray line profile case the tail of the probability distribution is not 
affected by the actual dislocation arrangement, but only by the average number of dislocations crossing the unit 
surface. To demonstrate this we took a set of 512 parallel edge dislocations with Burgers 
vectors parallel to the horizontal axis.
\begin{figure}[!ht]
\begin{center}
\includegraphics[width=0.3\textwidth]{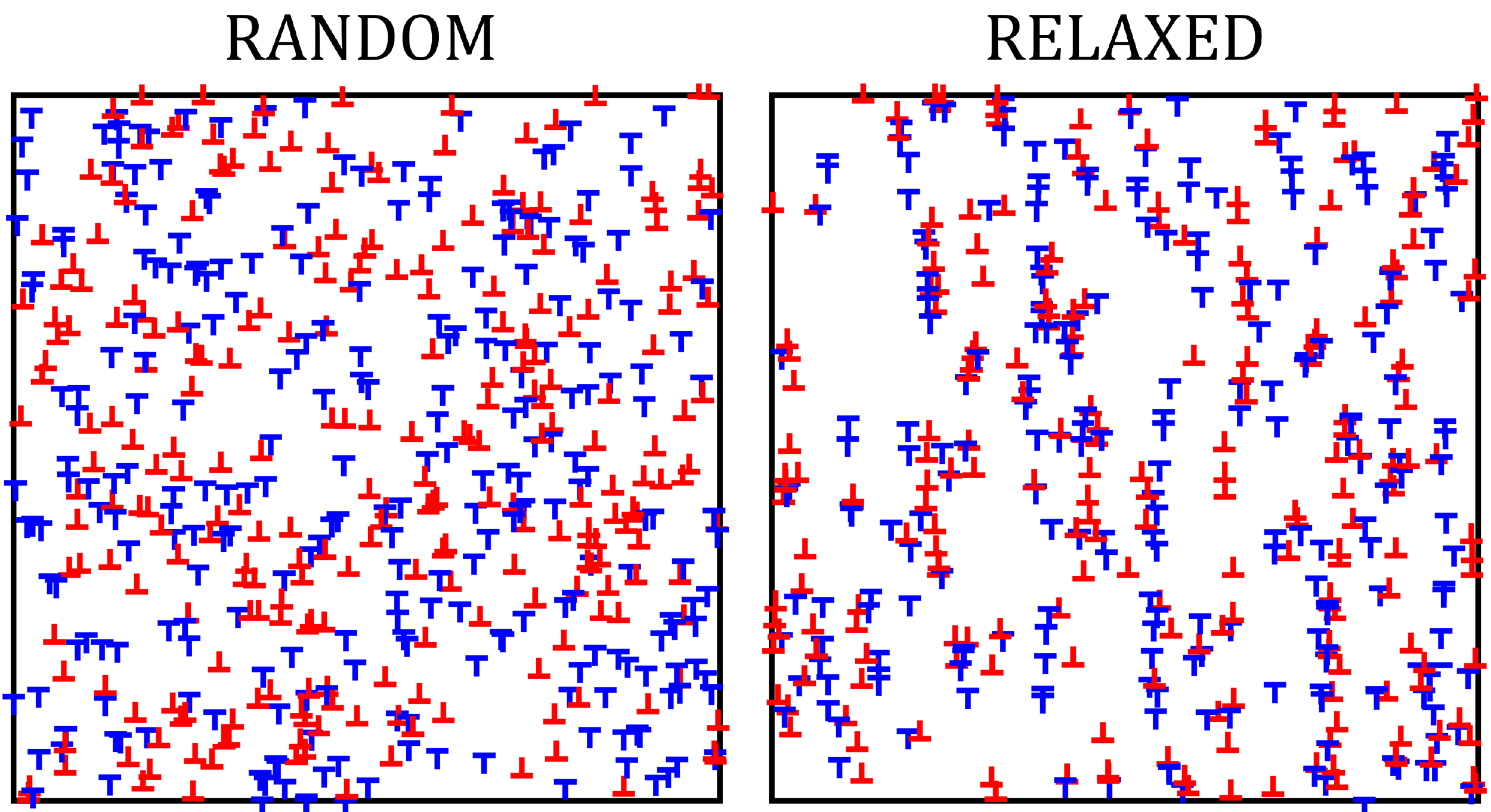} \\
\includegraphics[angle=0,scale=0.1]{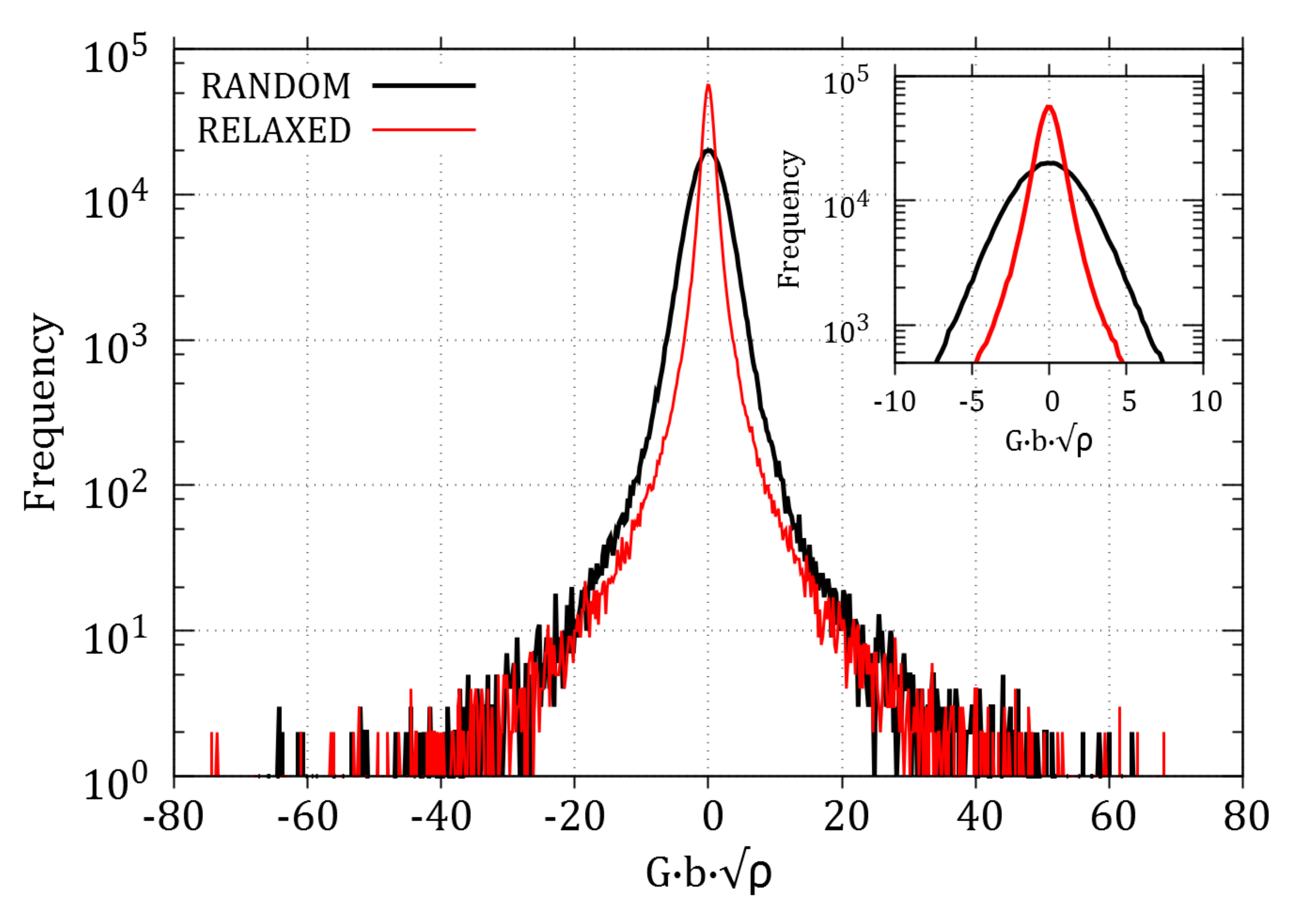}
\caption{\label{fig:dconf} Top boxes: Random and relaxed dislocation configuration. Bottom box:
Internal 
stress distributions obtained on the random (black curve) and the relaxed (red curve) 
configurations.  In the inset the central part of the distributions
are enlarged.}
\end{center}
\end{figure}
Initially the dislocations were placed randomly in a square box, then the system was relaxed with an 
over-dumped dynamics \cite{ispanovity2010submicron}. For the initial and the relaxed configurations 
(Fig. \ref{fig:dconf}/(top boxes)) the 
probability distribution of the shear stress was numerically determined by taking the stress values 
at $10^6$ randomly selected points. As seen in Fig. \ref{fig:dconf}/(bottom box) the tail of 
the 
distribution is not affected by 
the relaxation (in agreement with theoretical predictions \cite{groma1998,groma2004}), 
while the central region of $P(\sigma)$ becomes narrower in the relaxed state (inset in Fig. 
\ref{fig:dconf}/(bottom box)). 
It is important to note that for a completely random dislocation distribution the half width of $P(\sigma)$ tends to infinity with 
the logarithm of the system size, while for the relaxed configuration this divergence is canceled 
by 
dislocation-dislocation correlations \cite{groma1998}. 
So, due to stress screening caused by spatial correlations the distribution $P(\sigma)$ becomes 
independent from 
the size of the system \cite{groma1998}. Similarly to Bragg peak broadening the tail of $P(\sigma)$ is 
inverse cubic in the asymptotic regime. Hence its second order moment becomes linear in 
$\ln(\sigma)$ 
with a slope proportional to the average dislocation density. 
\begin{figure}[!ht]
\begin{center}
\includegraphics[angle=0,scale=0.12]{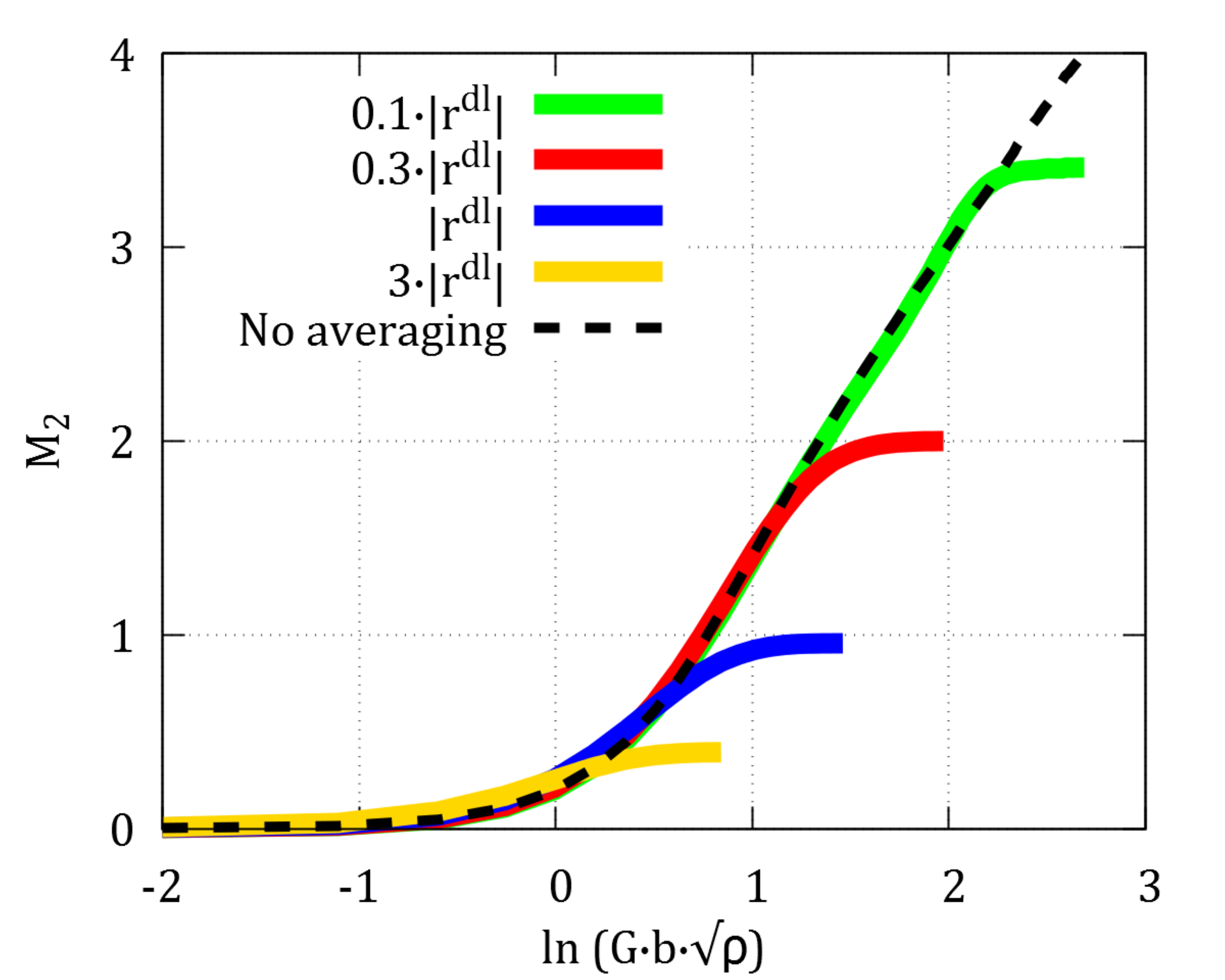} 
\caption{\label{fig:M_2} The $M_2(\sigma)$ vs. $\ln(\sigma)$ for 4 different averaging 
sizes and with no averaging,
corresponding to the relaxed configuration shown in Fig. \ref{fig:dconf}.}
\end{center}
\end{figure}

Due to the finite volume illuminated by the electrons in the SEM, a physically correct 
interpretation of experimental stress distributions requires averaging the theoretical 
distributions over the volume illuminated.  This introduces a cut-off in the inverse 
cubic decay of $P(\sigma)$. As a consequence the plot of $M_2$ versus $\ln(\sigma)$ deviates from 
the expected linear behavior 
as demonstrated in Fig. \ref {fig:M_2} showing the second order restricted moments corresponding 
to four "spatially averaged distributions" calculated for circles with diameters equal to 
$0.1 r^{dl}$, $0.3 r^{dl}$, $r^{dl}$, and $3 r^{dl}$,
where $r^{dl}$ is the average dislocation-dislocation distance. The curve with no 
applied averaging is also shown.
As expected the stress level at the cut-off is decreasing with increasing diameter or dislocation 
density. Therefore, during the evaluation 
of real data the cut-off introduced by the finite beam size should be considered in the analysis. Since the 
characteristic linear size of the illuminated volume (of about $10 \times 10 \times 50$ 
nm$^3$) \cite{yao2006,chen2011} 
can be of the same order of magnitude as the average dislocation-dislocation spacing in a heavily deformed metal  
(of  $\approx 30$ nm for a dislocation density of  $\approx 10^{15}$m$^{-2}$) the finite beam size could 
become a limiting factor for the application of the method.

To check the reliability of the EBSD method for dislocation density evaluation subsequent 
analyses were done by HR-EBSD and XRD on the same crystal  surfaces. Cu single crystals
of rotated Goss orientation $(011)[0\bar{1}1]$ were cut by electrical discharge machining into 
rectangular cuboid shapes 
and deformed by channel die compression up to strain levels of 6\% and 10\%. 
The compression was done along the normal direction (ND, parallel to the (011) plane 
normal), while the sample elongated along the longitudinal direction (LD)  
($[0\bar{1}1]$) and was held fixed by the channel walls along the transverse direction (TD).  
Before 
deformation and analysis by HR-EBSD and XRD the samples 
were electro-polished at 15 V for 3 minutes using the Struers D2 electrolyte. 
The procedure allowed obtaining Kikuchi patterns with the highest image quality.

\begin{figure}[!ht]
\begin{center}
\includegraphics[width=0.25\textwidth]{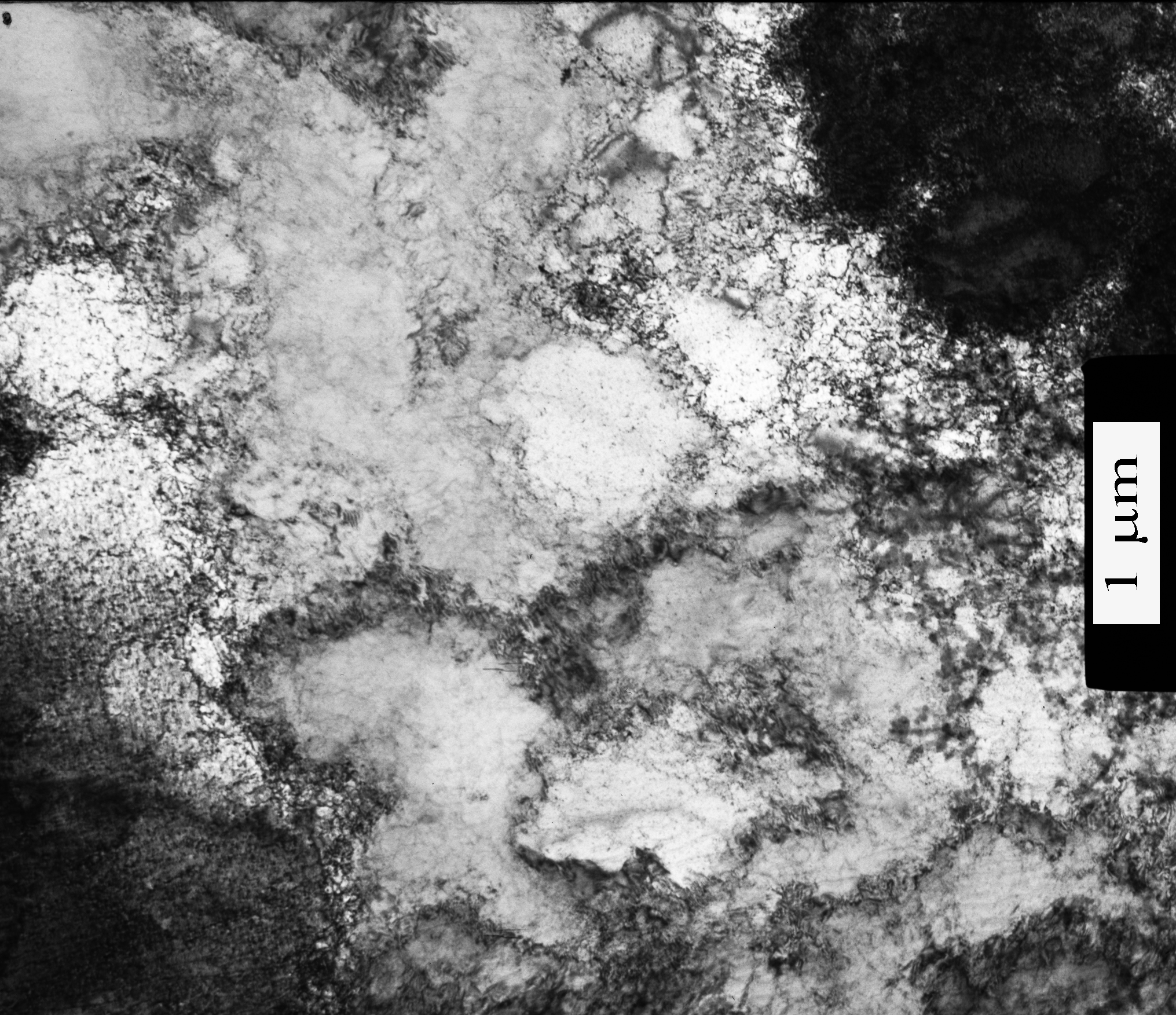} \\
\caption{\label{fig:TEM} TEM image showing the dislocation cell structure of the crystal deformed 
up to 6\% strain. }
\end{center}
\end{figure}

The rotated Goss orientation deforms homogeneously in channel die compression. 
According to the TEM image shown in Fig. \ref{fig:TEM} a well defined dislocation cell structure develops already at 6\% when the cell size is of 
about 1 $\mu$m. The samples were then characterized by XRD by measuring the 200 line profile on their TD surface. 
The measurements were done with Cu $K_{\alpha_1}$ radiation in a Panalytical MRD diffractometer  
equipped with a Bartels 
primary monochromator and a double bounce analyser, both made of Ge.

\begin{figure}[!ht]
\begin{center}
\includegraphics[scale=0.15]{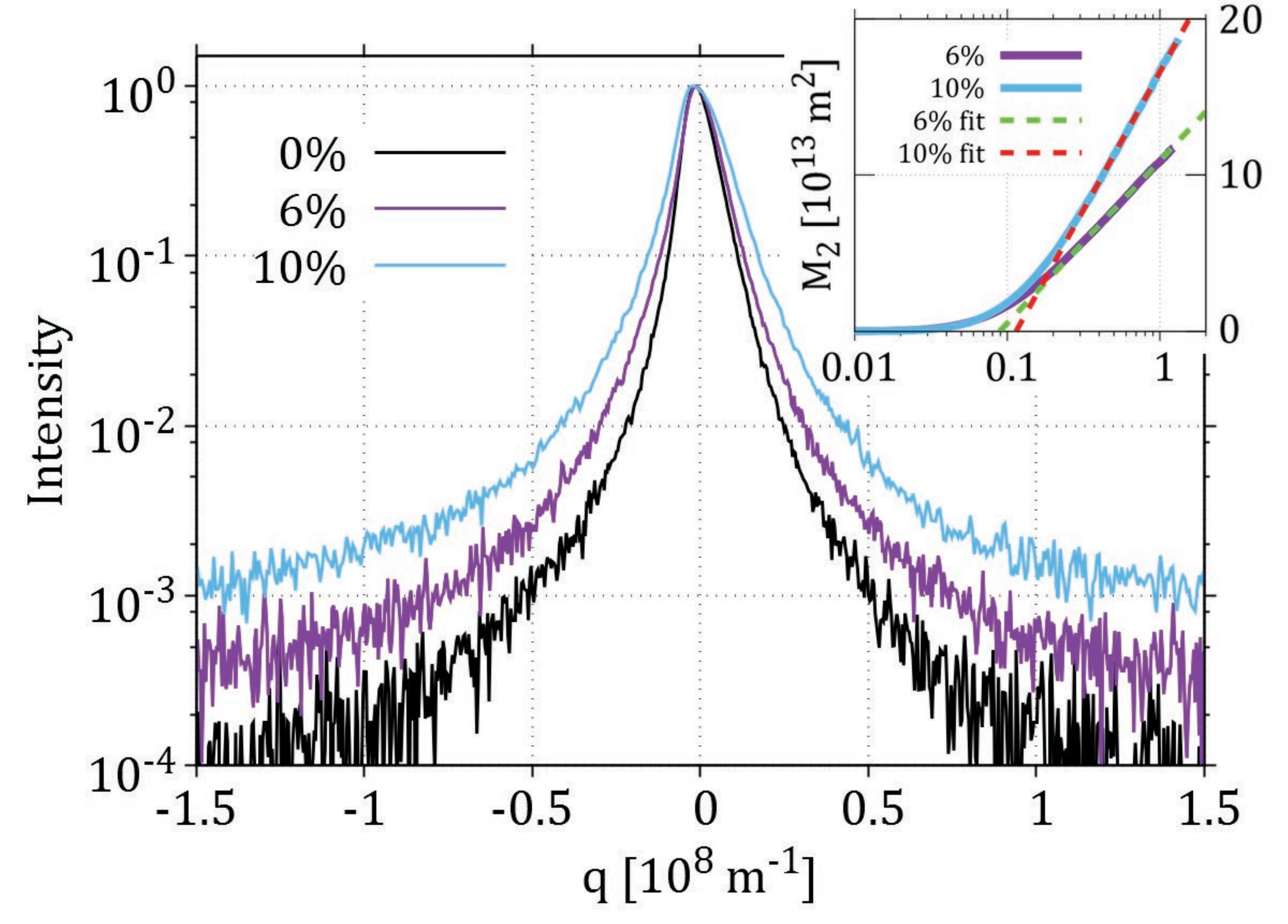}
\caption{\label{fig:xray} 200 X-ray Bragg peaks corresponding to 0\%, 6\%, and 10\% 
strain. 
Inset: The variance $M_2$ vs. $\ln(q)$ of the peaks measured on deformed samples. The straight 
lines are fits of the asymptotic regime.\vspace*{-0.3cm}}
\end{center}
\end{figure}

The 200 peaks and their variances $M_2(q)$ versus $\ln(q)$ are shown in Figs. 
\ref{fig:xray} and \ref{fig:xray}/(Inset), respectively. The dislocation density is 
directly obtained from the slope of the lines fitted to the asymptotic regime. The results given in 
the second column of Table 1 were obtained using 
a diffraction contrast factor $ C_g = 0.397$ corresponding to an equal dislocation population in each slip system.

\begin{table}[!ht] 
\begin{tabular}{|c|c|c|}\hline 
strain & $\rho_{XRD}$ & $\rho_{EBSD}$ \\
\hline 6\% & $7.3 \cdot 10^{14}$m$^{-2}$ & $2.3 \cdot 10^{14}$m$^{-2}$ \\
\hline 10\% & $1.2 \cdot 10^{15}$m$^{-2}$ & $1.3 \cdot 10^{15}$m$^{-2}$\\
\hline
\end{tabular} 
\centering \caption{Dislocation densities obtained by X-ray line profile analysis and HR-EBSD.} 
\label{table:xray} 
\end{table} 

The EBSD scans were done with a step size of 100 nm on a square grid covering 
an area about 25 $\mu$m $\times$ 30 $\mu$m. The backscattered Kikuchi patterns were recorded 
with a NordlysNano detector of 1344 $\times$ 1024 pixels. 
The acquisition was monitored with the AztecHKL software, which was also used to calculate the pattern centers 
necessary for performing the high-resolution evaluation. 
The stress at each measurement point was determined with the cross-correlation method of Kikuchi patterns 
developed by Wilkinson {\it et al.} \cite{britton2012}.
Since the scanned area is much larger than the characteristic size of the microstructure 
(dislocation cells with a size of about 1 $\mu$m, see Fig. \ref{fig:TEM}), the probability distribution of internal stresses 
can be considered a macroscopic quantity characterizing the structure.

\begin{figure}[!ht]
\begin{center}
\includegraphics[scale=0.17]{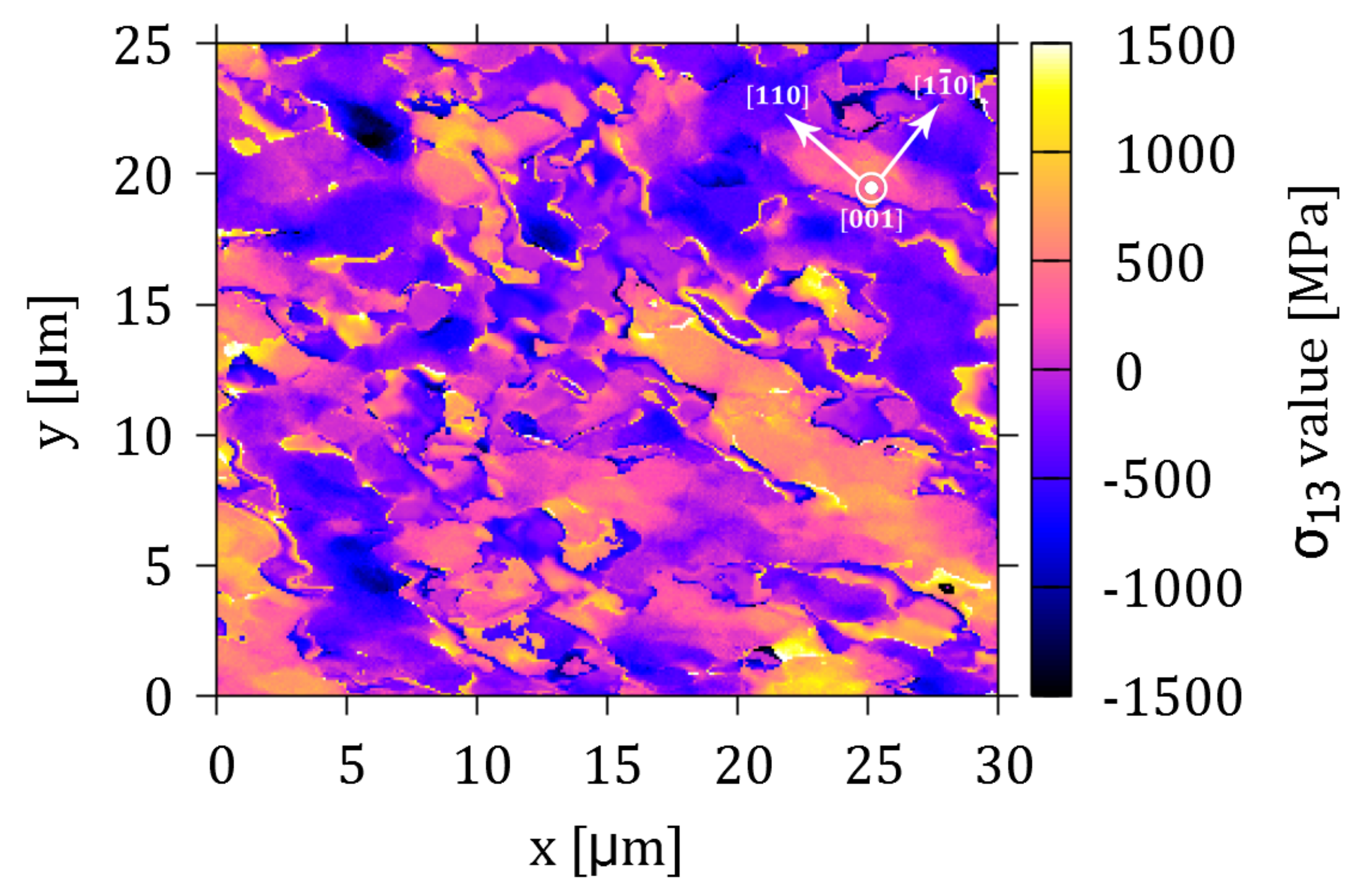} 
\caption{\label{fig:sigma} The $\sigma_{13}$ stress component map obtained by HR-EBSD 
on the Cu single crystal compressed to 6\% strain. The stress levels indicated are relative 
values to the stress level at the center of the area scanned.\vspace*{-0.3cm}}
\end{center}
\end{figure}

The $\sigma_{13}$ stress component map obtained on the sample with 6\% strain is plotted in Fig. 
\ref{fig:sigma}. In agreement with the TEM image (Fig: \ref{fig:TEM}), the cell 
structure with typical cell size of 1$\mu$m is seem with long range internal stress develops in the 
cell interiors \cite{mughrabi}. A detailed analysis of the long range stress is out of the scope of 
the present paper, it will be published elsewhere.

\begin{figure}[!ht]
\begin{center}
\includegraphics[scale=0.15]{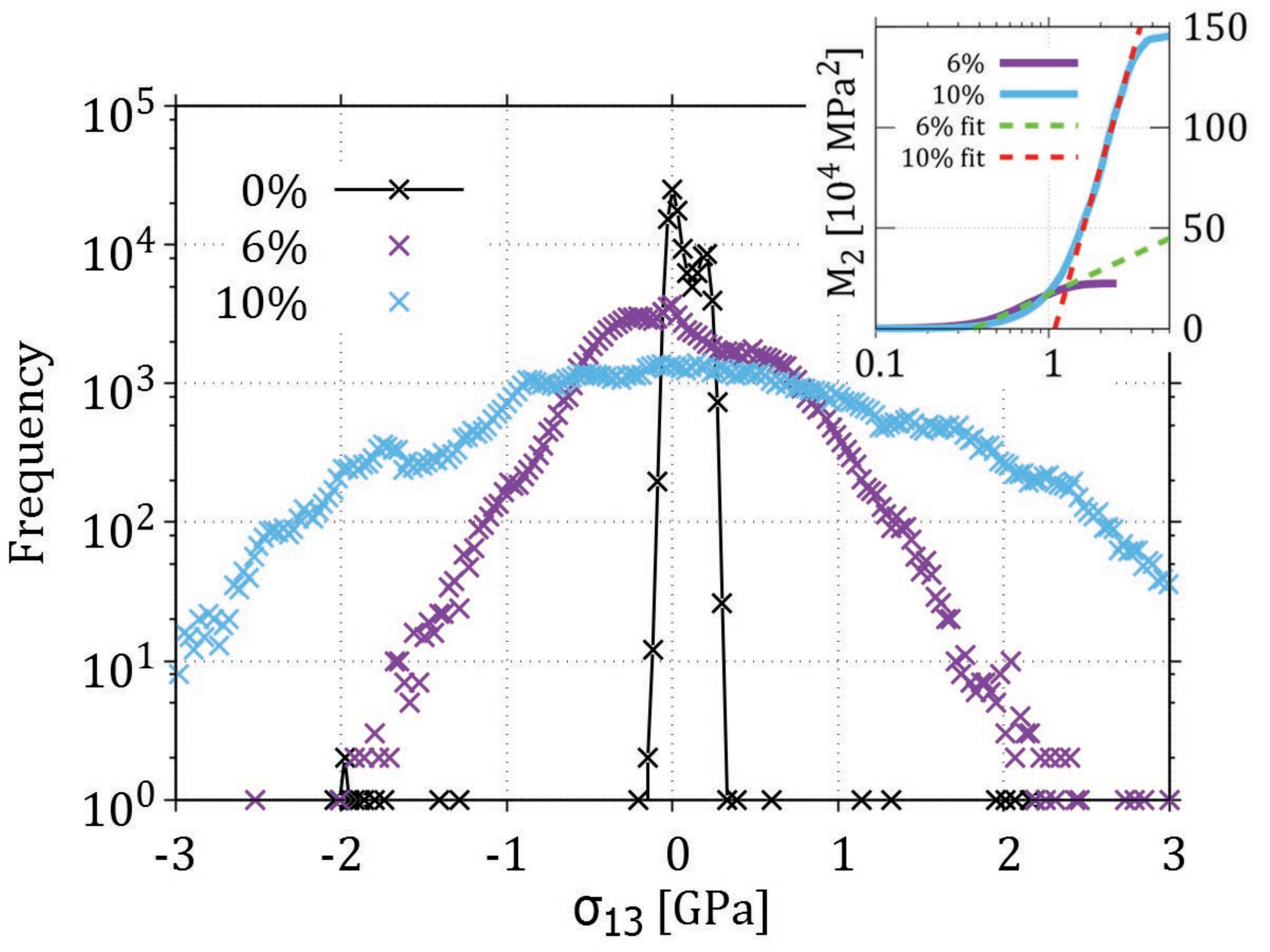} 
\caption{\label{fig:psigma} a) The probability distribution of the $\sigma_{13}$ stress 
component at strains of 0\%, 6\% and 10\% . 
Inset: The corresponding variances $M_2$ versus $\ln(q)$ for the deformed samples, with the 
straight lines fitted in the asymptotic regime. \vspace*{-0.3cm}}
\end{center}
\end{figure}

The probability distributions of the $\sigma_{13}$ stress component characterizing the undeformed 
and deformed samples are plotted in Fig. \ref{fig:psigma}. $P(\sigma_{13})$ is very 
narrow for the undeformed 
sample 
and it broadens with increasing deformation. 
Remarkably, the tails of $P(\sigma_{13})$ extend outside values as large as $\pm 1$ GPa. Similar 
behavior was first reported by Wilkinson {\it et al.} on deformed polycrystalline Cu and ferritic 
steel \cite{wilkinson2014}. 
Since one can clearly identify a linear regime on the $M_2$ versus $\ln(\sigma_{13})$ plots 
(Fig. \ref{fig:psigma}/(Inset)) the broadening of $P(\sigma_{13})$ can only be caused by the 
presence of 
dislocations (other type of stress source would generate different decay in the tail part \cite{groma1998}). 
This reasoning is also supported by the rather narrow $P(\sigma_{13})$ distribution corresponding to
the undeformed sample. For stress values larger than about 2 GPa the second 
order restricted moments clearly deviate from the linear dependence in $\ln(\sigma_{13})$. 
As discussed above this is the consequence of the measuring setup and related to the unavoidable averaging 
over the volume illuminated by the electron beam. Nevertheless, the linear regime can be well 
identified on the plots presented. (The cut-off is certainly absent on the variance of the X-ray 
peaks.) According to Fig. \ref {fig:M_2} the linear region disappears when the size of  the 
averaging zone equals the mean dislocation-dislocation spacing. This imposes an instrumental limit 
in the application of the method for 
heavily deformed samples.

According to Eq. (\ref{eq:P}) the slope of the line fitted in the asymptotic regime is proportional 
to the total dislocation density. Its determination requires the knowledge of the 
{\it {stress contrast factor}} $C_{\sigma}$ in Eq. (\ref{eq:P}), 
which can be calculated according to Refs. \cite{groma1998,groma2004}. 
For the stress component $\sigma_{ij}$ considered in the analysis one has to evaluate the integral:
\begin{eqnarray}
 C_{ij}=\frac{1}{G^2 b^2}\int_0^{2\pi} \left[r\sigma^{\rm ind}_{ij}(r,\varphi)\right]^2 {\rm d} 
\varphi \label{eq:c}
\end{eqnarray}
where $\sigma^{\rm ind}_{ij}$ is the stress generated by a dislocation with a given line direction 
$\vec{l}$ and Burgers vector $\vec{b}$ in the $xy$ plane of the coordinate system in which the stress tensor 
is calculated during the evaluation of the Kikuchi patterns. In Eq. (\ref{eq:c}) $(r,\varphi)$ 
denotes the polar 
coordinates in the $xy$ plane. (Due to the $1/r$ type of decay of the stress field generated by a 
dislocation, the integral is independent from $r$.) Since in anisotropic materials the stress field 
of a straight dislocation cannot be always given in a closed analytical form \cite{steeds73} $C_{\sigma}$ can 
only be calculated numerically. Moreover, since in most cases dislocations of different types and 
line directions can exist in the same structure, one has to calculate the appropriate weighted average of  
$C_{\sigma}$ corresponding to given $\vec{b}$ and $\vec{l}$. This issue is out of the scope of this 
paper. For simplicity we use the value corresponding to an edge dislocation with line direction 
perpendicular to the sample surface. In this case $C_{\sigma}=1/(8 \pi (1-\nu)^2)$ where $\nu$ is the 
Poisson number \cite{groma1998}.

The dislocation densities of the deformed samples are summarized in the third column of Table 1. 
The $\rho_{EBSD}$ values given correspond to the average values obtained from the stress components 
$\sigma_{13}$, $\sigma_{22}$, and $\sigma_{23}$. (Due to the deformation geometry 
applied the other two stress components  $\sigma_{11}$, and  $\sigma_{12}$ are much smaller with 
much larger error, so they were not taken into account.)
By comparing them to the values obtained from XRD (column 2) one can conclude that there is a 
good agreement between the results of the two methods. At 10\% strain the difference is
within a few \% of relative error, while at 6\% strain the HR-EBSD gave smaller $\rho$ than the XRD 
by about a factor of 3. 
The last difference can be attributed first of all to the influence of the larger dislocation cell 
size at 6\% strain resulting that the volume scanned during the EBSD measurement may not large 
enough to give a representative mean value for the dislocation density.  
Another reason for the difference can be a change in the main dislocation character 
and the population of different slip systems with increasing strain. It seems to be that the 
$C_{\sigma}$ used is not really relevant for the 6\% strain case. The issue requires further 
detailed investigations.

It is remarkable, however, that the 
assumption considering edge dislocations only gave good agreement with XRD results at 10\% strain. 
This emphasizes the strong physical basis of the evaluation method proposed. For more accurate 
dislocation density values a TEM analysis of prevailing dislocation types and line vectors is 
inevitable both for the XRD and HR-EBSD methods  

Summing up: HR-EBSD was traditionally used to determine the \emph{geometrically 
necessary dislocation 
density}\cite{randman}. With the analysis of the tail of the stress 
probability-distribution function 
obtained from HR-EBSD the stored, \emph{total dislocation density} present in the sample can also be 
determined. This opens new perspectives for the application of EBSD in determining
\emph{mesoscale} parameters in a heterogeneous sample, such as a polycrystal.

The authors acknowledge insightful discussion with Prof. Claire Maurice (Ecole Nationale 
Sup\'erieure des Mines, Saint-Etienne, France).  
This work was supported by the French-Hungarian collaboration BALATON, the Hungarian Scientific 
Research Fund (OTKA) under contract numbers K-105335 and PD-105256, and the European Commission 
under grant agreement No. CIG-321842. PDI was supported by the J\'anos Bolyai Scholarship of the 
Hungarian Academy of Sciences. 


\end{document}